\documentclass[onecolumn,showpacs,prl,superscriptaddress]{revtex4}
\usepackage{graphicx}
\usepackage{color}
\usepackage{amssymb,amsfonts,amsmath}
\usepackage{natbib}

\begin{document}

\title{Fluctuating hydrodynamics approximation of the stochastic Cowan-Wilson model}

\author{Clement Zankoc}
 \affiliation{Dipartimento di Fisica e Astronomia, CSDC and INFN, Universit\`{a} degli Studi di Firenze, via G. Sansone 1, 50019 Sesto Fiorentino, Italia}
\author{Tommaso Biancalani}  \affiliation{Physics of Living Systems, Department of Physics, Massachusetts Institute of Technology, Cambridge, Massachusetts, United States of America}
\author{Duccio Fanelli}  \affiliation{Dipartimento di Fisica e Astronomia, CSDC and INFN, Universit\`{a} degli Studi di Firenze, via G. Sansone 1, 50019 Sesto Fiorentino, Italia}
\author{Roberto Livi}  \affiliation{Dipartimento di Fisica e Astronomia, CSDC and INFN, Universit\`{a} degli Studi di Firenze, via G. Sansone 1, 50019 Sesto Fiorentino, Italia}

\begin{abstract}
We consider a stochastic version of the Wilson-Cowan model which accommodates for discrete populations of excitatory and inhibitory neurons. The model assumes a finite carrying capacity with the two populations being constant in size. The master equation that governs the dynamics of the stochastic model is analyzed by an expansion in powers of the inverse population size, yielding a coupled pair of non-linear Langevin equations with multiplicative noise. Gillespie simulations show the validity of the obtained approximation, for the parameter region where the system exhibits dynamical bistability. We report analytical progress by silencing the retroaction of the activators on the inhibitors, while still assigning the parameters so to fall in the region of deterministic bistability for the excitatory species.  The proposed approach forms the basis of a perturbative generalization which applies to the case where a modest degree of coupling is restored.
 \end{abstract}

\pacs{89.75.Hc 89.75.Kd 89.75.Fb}

\maketitle


\section{Introduction}
\label{sec1}

Neural models aim at elucidating the grounding architecture that regulates information processing in biological nervous systems \cite{Kandel}. The level of abstraction that is accommodated for in the chosen mathematical scheme reflects the  specific topic under investigations.  Detailed single neurons models can be devised, which account for the sharp changes in the action potential. The number of neurons in the human cortex is extraordinarily large and for many practical aspects a bottom-up approach that moves from a detailed representation of each individual unit is virtually impracticable. The physiology of neurons is also quite intricate. A single neuron receives input from thousand of axon terminals coming from other neurons, but the inherent degree of redundancy contributes to drastically reduce the effective connections to a limited subset. Almost $80 \%$ of the neurons are excitatory, whereas the remaining $20 \%$ belong to the inhibitory type. Irrespectively of their intimate characteristics, neurons can be pictured as complex switching devices which organize into two macroscopic families of homologous constituents,  the activators and inhibitors. This enables in turn to climb the hierarchy of complexity, and elaborate, to a suitable level of coarse graining, on the interlaced dynamics of homogenous populations of binary neurons.

The celebrated Wilson-Cowan (WC) model \cite{WC1,WC2} and its further extensions provide an interesting arena to shed light onto the complex dynamics of a multi-species neuronal system. The model displays elementary limit cycle behavior, mimicking in silico neural oscillations. Stimulus-dependent evoked responses are also predicted, as well as the existence of multiple stable states. The original WC model is deterministic in nature and the average concentration of active/inactive excitatory/inhibitory neurons appear to be self-consistently ruled by a pair of coupled ordinary differential equations. Spatially extended versions of the model can be straightforwardly designed which in turn amount to operate with partial differential equations \cite{Kilpatrick}.

Deterministic systems can be analyzed by resorting to a gallery of well-developed tools from the theory of dynamical systems and involving concepts such as trajectories in phase space and attractors of various kinds \cite{strogatz, risken}. Stochastic systems exhibit an equivalently rich plethora of attributes, ranging from stochastic fluctuations about the deterministic trajectories, rare events which prompt the transitions from one attractor of the deterministic dynamics to another, stationary probability distributions, to mention a few \cite{gardiner}.  The stochasticity can be externally imposed and hence impact on the deterministic scenario, as an exogenous source of disturbance. More intriguing is instead the role played by intrinsic perturbation, stemming from the intimate discreteness of the system under scrutiny. Individual-based effects should be unavoidably accommodated for in any sensible model of natural phenomena and yield an endogenous stochastic contribution, which lacks instead under the idealized continuum representation. At first sight it might appear surprising that stochastic effects are important when the interacting population consist of a large number of individual constituents. Demographic fluctuations can however amplify through resonant effects \cite{bartlett} and drive the spontaneous emergence of collective macroscopic patterns, both in time~\cite{mckanePRL, dipatti} and in space~\cite{lugo, butler, biancalani, deanna, woolley}, marking a clear distinction between stochastic and deterministic viewpoints.

Endogenous noise is certainly relevant to neural systems \cite{goldenfeld}. Channel noise arising from the variability in the opening and closing of a finite number of ion channels, is a microscopic source of intrinsic disturbance. At the population level, the instantaneous photograph of the system is drawn by looking at the number of active neurons, which sits on a given patch. State transition can be assigned to monitor the evolution of the inspected system via a suitably defined master equation, the mean field deterministic picture being eventually recovered in the thermodynamic limit. Following these lines of reasoning, the WC model has been recently revised under the generalized stochastic angle, so to cast the accent on the specific traits which ultimately emanates from the endogenous component of the noise \cite{bressloff}. Our paper is positioned in this context \cite{Goychuck,Negahbani,Cowan2016}. We will in particular reformulate the WC dynamics as a simple birth and death stochastic process, with non linear 
 transition rates. The obtained master equation will be expanded in powers of the inverse population size so to yield an approximate stochastic description for the density of the interacting species. More specifically, the fluctuating hydrodynamics for the stochastic WC model is represented by a pair of non linear Langevin equations, with multiplicative noise. These latter are shown to provide an adequate description of the original stochastic process, the analysis being specifically targeted to the domain of dynamical bistability.  To gain further insight we will confine ourselves to inspecting the special setting where the bistability solely interests one the two species, the excitatory neurons, while the concentration of the inhibitors converges to a stationary stable fixed point, in the limit of diverging population. Working in this setting, one can effectively reduce the system to just one Langevin equation for the excitatory species: an additional contribution appears in the expression for the multiplicative noise, which exemplifies the action exerted by the inhibitory neuron. The population of inhibitors acts as a source of intrinsic noise, shaking the discrete ensemble of excitators from the {\it inside}. Interestingly, the perturbation magnifies as $\gamma$, the relative ratio of inhibitory versus excitatory neurons, gets reduced. For actual physiological conditions ($\gamma \simeq 1/4$), the mechanism here identified is hence potentially relevant. Finite size noise and its inhibitory based component causes an extension of the region deputed to the excitatory bistability, as compared to deterministic predictions.  This contribute in turn to broaden the  plasticity of the system and elevate its degree of inherent flexibility.  The analytical tools developed in the early part of the paper are subsequently generalized to the relevant setting where both species can undergo bistable dynamics. A recursive perturbative scheme targeted to such a general scenario is successfully implemented and tested versus direct numerical simulation.

The paper is organized as follows. In the next Section we will introduce the stochastic version of the Wilson-Cowan model and carry out the analysis that yields to the fluctuating hydrodynamical picture. In Section 3, we will silence the retroaction of the activators on the inhibitors, and assign the parameters so to fall in the region of deterministic bistability for the excitatory species.  By linearizing around the fixed point of the inhibitors, one can reduce the overall dynamics to a non-linear Langevin equation for the density of activators: the multiplicative noise incorporates the effect which stems from the population of inhibitors. Working in this framework, one can analytically determine the probability distribution for both species and quantitatively elaborate on their mutualist interdependence. Noise-induced bistability is uncovered, the effect being progressively more evident the smaller the population ratio $\gamma$. Building on these  premises, 
 in Section 4 we generalize the study to the setting where the retroaction  of the activators is restored. Finally in Section 5 we sum up and draw our conclusions.

\section{A stochastic version of the Wilson-Cowan model}
\label{sec2}

The Wilson-Cowan (WC) model describes the self-consistent interactions between distinct populations of excitatory and inhibitory neurons. We shall hereafter provide a plausible microscopic formulation of the dynamics that eventually converges to the sought WC model in the deterministic limit. As mentioned earlier, by providing a microscopic characterization of the scrutinized dynamics we aim at shedding light onto the role played by finite size fluctuations. In particular, working under the Kramers-Moyal approximation we will derive an effective description of the stochastic model in terms of fluctuating hydrodynamical equations. The model deals with a spatially homogeneous distribution of the interacting entities.  Extensions of the proposed scheme which assume the constituents dislocated on the nodes of a complex graph, so to probe the heterogeneity in the emerging distribution, are certainly conceivable.

In the following, individual excitatory and inhibitory neurons will be respectively identified with the symbols $X$ and $Y$. Both excitatory and inhibitory neurons can manifest into two distinct states: activated ($a$) or quiescent ($q$). To identify to which subpopulation a neuron belongs, the quantities $X$ and $Y$ will be decorated with the index $a$ or $q$. More concretely,
we will denote by $X_a$ (resp. $Y_a$) an activated neuron of the excitatory type (resp. inhibitory), whereas $X_q$ (resp. $Y_q$)  identifies a quiescent excitatory  (resp. inhibitory) neuron.
The stochastic dynamics of the system is ruled by a minimal set of birth and death reactions, as summarized in the following:

\begin{eqnarray}
\label{chemical_eqs}
X_q&&\overset{f_E[s_E]}{\longrightarrow} X_a 
\nonumber \\
X_a&&\overset{\alpha_E}{\longrightarrow} X_q 
\nonumber \\
Y_q&&\overset{f_I[s_I]}{\longrightarrow} Y_a 
\nonumber \\
Y_a&&\overset{\alpha_I}{\longrightarrow} Y_q
\end{eqnarray}

where $\alpha_{E,I}$ are constant rate function; $f_{E,I}(\cdot)$ are non linear firing rate, sigmoid in shape, function of the currents  $s_{E,I}$, and bound to the interval $[0,1]$. Before completing the description of the elements that define the stochastic model, we remark that, by definition, the number of excitatory (resp. inhibitory) neurons, $N$ (resp. $M$), is an invariant quantity. Label with $n_{X_a}$ the number of neurons of the type $X_a$, and with $n_{X_q}$ the number of elements $X_q$. Clearly, by virtue of equations (\ref{chemical_eqs}), $n_{X_a}+n_{X_q}=N$. Similarly, and with an obvious meaning of the involved notation $n_{Y_a}+n_{Y_q}=M$. Simply stated, the conservation laws that we have here identified enable one to reduce to a total of two the number of independent variables that uniquely lay out the dynamics of the system. We then set:

\begin{subequations}
\begin{align}
s_E=\omega_{EE}  \frac{n_{X_a}}{N} - w_{EI} \frac{n_{Y_a}}{M} + h_E \\
s_I=\omega_{IE} \frac{n_{X_a}}{N} - w_{II} \frac{n_{Y_a}}{M}  + h_I
\end{align}
\end{subequations}

where the weights $\omega_{kl}$ are positive defined. The positive constants $h_E$ and $h_I$ encode the interaction of the examined families of neurons with the surrounding environment. The specific form of the sigmoid function $f_{k}(s_k)$,  with $k=E,I$ is not essential for the forthcoming discussion. To proceed in the analysis we introduce the number density
$x=n_{X_a}/N$ and $y=n_{Y_a}/M$ . We further label with $P(x,y,t)$ the probability of seeing the system in the state photographed by $(x,y)$, at time $t$. The chemical equations introduced above define a Markov process, that is ruled by the following Master equation:

\begin{eqnarray*}
\frac{\partial P(x,y,t)}{\partial t}  & = & [ - T_1(x+\frac{1}{N},y|x,y)P(x,y,t) -T_2(x-\frac{1}{N},y|x,y)P(x,y,t)\\
& &  -T_3(x,y+\frac{1}{M}|x,y)P(x,y,t) -T_4(x,y-\frac{1}{M}|x,y)P(x,y,t) \\
& &  +T_5(x,y|x-\frac{1}{N},y)P(x-\frac{1}{N},y,t) +T_6(x,y|x+\frac{1}{N},y)P(x+\frac{1}{N},y,t) \\
& &  +T_7(x,y|x,y-\frac{1}{M})P(x,y-\frac{1}{M},t)+T_8(x,y|x,y+\frac{1}{M})P(x,y+\frac{1}{M},t)]
\end{eqnarray*}

where $T_i( \cdot | \cdot)$ are the transition rates. Before making explicit the definition of the transition rates, we write the Master equation in a compact form involving the so called step operators $\epsilon^{\pm}_{x}$ and $\epsilon^{\pm}_{y}$ defined as:

\begin{subequations}
\begin{align}
\epsilon^{\pm}_{x}f(x)=f(x\pm \frac{1}{N}) \\  
\epsilon^{\pm}_{y}f(y)=f(y\pm \frac{1}{M})
\end{align}
\end{subequations}
 
Making use of the above operators, one can re-write the Master equation in the form:

\begin{eqnarray}
\label{ME_eps}
\frac{\partial P(x,y,t)}{\partial t}  & = & [(\epsilon^{-}_{x}-1)T_1 + (\epsilon^{+}_{x}-1)T_2
\nonumber  \\
& & (\epsilon^{-}_{y}-1)T_3+(\epsilon^{+}_{y}-1)T_4]P(x,y,t)
\end{eqnarray}

where:

\begin{subequations}
\begin{align}
T_1(x+\frac{1}{N},y|x,y)=(1-x)f_E(x,y) \\
T_2(x-\frac{1}{N},y|x,y)=\alpha_E  x \\
T_3(x,y+\frac{1}{M}|x,y)=(1-y)f_I(x,y) \\
T_4(x,y-\frac{1}{M}|x,y)=\alpha_I  y 
\end{align}
\end{subequations}

The Master equation (\ref{ME_eps}) provides an exact description of the Markov dynamics. In particular, it contains
information on both the ideal mean field dynamics (formally recovered in the limit of diverging system size)
and the associated finite size corrections. To elaborate on those aspects, we seek to approximate the exact
Master equation, via the Kramers-Moyal perturbative recipe \cite{vk,rogers,biancalani1,biancalani2}. To this end we
Taylor expand the step operators (assuming $1/N$ and $1/M$ small) and get:
\begin{eqnarray}
\epsilon^{\pm}_{x}f(x)=f(x\pm \frac{1}{N})\approx (1\pm \frac{1}{N}\frac{\partial }{\partial x}+\frac{1}{2N^2}\frac{\partial^2}{\partial x^2}+\dots)f(x) \\
\epsilon^{\pm}_{y}f(y)=f(y\pm \frac{1}{M})\approx (1\pm \frac{1}{M}\frac{\partial }{\partial y}+\frac{1}{2M^2}\frac{\partial^2}{\partial y^2}+\dots)f(y)
\end{eqnarray}

Then, inserting in the Master equation, one eventually obtains:
\begin{eqnarray*}
\frac{\partial P(x,y,t)}{\partial t}  & = & [(-\frac{1}{N}\frac{\partial }{\partial x}+\frac{1}{2N^2}\frac{\partial^2}{\partial x^2})T_1 + (\frac{1}{N}\frac{\partial }{\partial x}+\frac{1}{2N^2}\frac{\partial^2}{\partial x^2})T_2 \\
& & (-\frac{1}{M}\frac{\partial }{\partial y}+\frac{1}{2M^2}\frac{\partial^2}{\partial y^2})T_3+(\frac{1}{M}\frac{\partial }{\partial y}+\frac{1}{2M^2}\frac{\partial^2}{\partial y^2})T_4]P(x,y,t)
\end{eqnarray*}
and performing the obvious time rescaling $\tau=Nt$:
\begin{eqnarray*}
\frac{\partial P(x,y,t)}{\partial \tau}  & = & [(-\frac{\partial }{\partial x}+\frac{1}{2N}\frac{\partial^2}{\partial x^2})T_1 + (\frac{\partial }{\partial x}+\frac{1}{2N}\frac{\partial^2}{\partial x^2})T_2 \\
& & (-\frac{1}{\gamma}\frac{\partial }{\partial y}+\frac{1}{\gamma^2}\frac{1}{2N}\frac{\partial^2}{\partial y^2})T_3+(\frac{1}{\gamma}\frac{\partial }{\partial y}+\frac{1}{\gamma^2}\frac{1}{2N}\frac{\partial^2}{\partial y^2})T_4]P(x,y,t)
\end{eqnarray*}

where $\gamma=M/N$ measures the relative population of inhibitors over activators.  This is a two dimensional Fokker-Planck equation, which we shall write in a more transparent form by introducing the vector ${\bf x}=(x,y)$. In formulae:

\begin{equation}
\frac{\partial P({\bf x},t)}{\partial \tau} = -\sum_i\left[\frac{\partial}{\partial x_i}A_i({\bf x})P(\bf{x},t) \right] + \frac{1}{2 N}\sum_{i,j}\frac{\partial}{\partial x_i}\frac{\partial}{\partial x_j} \left[ B_{ij}({\bf x})P({\bf x},t) \right]
\end{equation}
where :
\[ A=  \left[\begin{array}{c}
T_1-T_2 \\
\frac{1}{\gamma}(T_3-T_4)
\end{array}\right] \]
and
\[ B=  \left[\begin{array}{cc}
T_1+T_2 & 0 \\
0 & \frac{1}{\gamma^2}(T_3+T_4)
\end{array}\right] \]

The above Fokker-Planck equation is equivalent to the following non linear Langevin equations \cite{risken}:
\begin{eqnarray}
\dot{x} &=&-(T_2-T_1)+\frac{1}{\sqrt{N}}\xi_1 \\
\dot{y}&=&-\frac{1}{\gamma}(T_3-T_4)+\frac{1}{\sqrt{N}}\xi_2
\label{lang0}
\end{eqnarray}
where the noise have zero mean and $<\xi_i (t) \xi_j (t')> = B_{ij} \delta(t-t')$, with $i,j=1,2$. Notice that the factor $\gamma$ has been incorporated in the definition of  matrix $B$. For this reason the noise term in the second of equations  (\ref{lang0}) scales as $1/\sqrt{N}$.
We now make the transformation $\xi_i=\sum^2_{j=1}G_{ij}\eta_j$, where $G$ is such that $B=GG^{T}$ and $<\eta_i (t) \eta_j (t')> = \delta_{ij} \delta(t-t')$.
In our case :
\[ G=  \left[\begin{array}{cc}
\sqrt{T_1+T_2} & 0 \\
0 & \frac{1}{\gamma}\sqrt{T_3+T_4}
\end{array}\right] \]

A straightforward calculation yields:

\begin{eqnarray}
\label{mult_lang}
\dot{x}&=&-\alpha_E x +(1-x)f_E(x,y)+\frac{1}{\sqrt{N}}\left(\sqrt{\alpha_Ex_1+(1-x)f_E}\right)\eta_1 \\
\dot{y}&=&\frac{1}{\gamma}\left(-\alpha_I y +(1-y)f_I(x,y)\right)+\frac{1}{\sqrt{\gamma N}}\left(\sqrt{\alpha_I y+(1-y)f_I}\right)\eta_2 \nonumber
\end{eqnarray}

\begin{figure}
 \centering
 \includegraphics[scale=0.37]{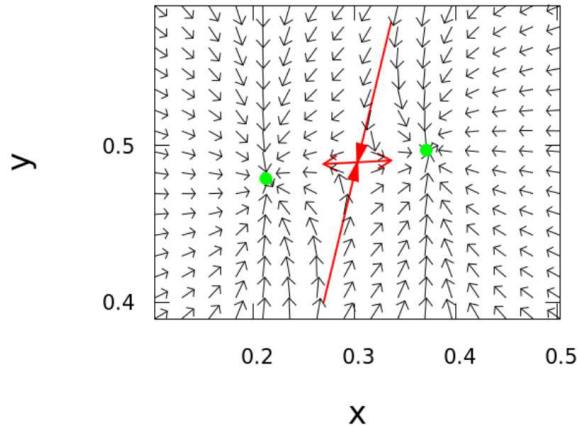}
 \caption{The phase space of the deterministic WC model is displayed. The (green) filled circles stands for the stable fixed points. Large (red) arrows represent the
 stable and unstable manifolds of the saddle node (not displayed). Tiny (black) arrows provide a quantitative description of the velocity vector field of the model in its deterministic limit.
 Here, $f_{E,I}=f^{(1)}_{E,I}+f^{(2)}_{E,I} \tanh (\beta_{E,I} s_{E,I})$  where $\beta$ is the gain parameter. The quantities $f^{(1)}_{E,I}$ and  $f^{(2)}_{E,I}$ enters the definition of the sigmoid
 non linear function and allow for a swift control of the location of the fixed points. The offset $f^{(1)}_{E,I}$ sets in particular the degree of
 residual activity when $s_{E,I}=0$. The parameters here employed read $\omega_{ii}=1$, $\omega_{IE}=0.5$, $\omega_{ei}=2$, $\omega_{ee}=7.2$
 $\alpha_E=1.5$, $\alpha_1=0.4$, $h_E=-1.2$, $h_I=0.1$, $N=300$, $\gamma=0.25$, $\beta_E=3.7$, $f^{(1)}_{E}=0.25$, $f^{(1)}_{E}=0.65$, $\beta_I=1$,
 $f^{(1)}_{I}=0.5$, $f^{(1)}_{I}=0.5$.}
 \label{fig1}
 \end{figure}

\begin{figure}[!h]
 \includegraphics[scale=0.4]{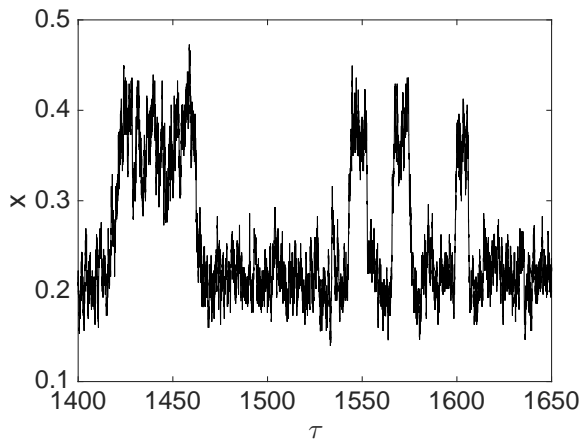}
 \caption{The number density $x=n_{X_a}/N$ is plotted versus time. The time series is obtained as an individual realization of the stochastic WC model, integrated via the Gillespie Monte Carlo scheme. The parameters are set as in Figure  \ref{fig1}.}
 \label{fig2}
 \end{figure}

We have hence obtained a system of non linear Langevin equations that constitutes the fluctuating hydrodynamics approximation of the exact Markov dynamics. Notice that the noise is multiplicative: the amplitude of the stochastic perturbation is self-consistently adjusted, as function of the density of the simultaneously evolving species. Performing the thermodynamics limit
$N \rightarrow \infty$ (at constant $\gamma$), the noise term fades away and one recovers a slightly modified version of the classical WC model.
As compared to the original WC formulation, the first (resp. second) equation of (\ref{mult_lang}) displays an additional factor $(1-x)$ (resp. $(1-y)$), which multiplies the sigmoid function $f_E$ (resp. $f_I$ ). This term reflects the finite carrying capacity of the system, as imposed at the microscopic level: working under diluted condition, i.e. assuming $x,y \ll 1$ one readily converge to the WC model in its primitive conception \cite{goldenfeld}. As already mentioned the deterministic WC model presents a large plethora of interesting dynamical behaviors. Depending on the assigned parameters, the system can evolve toward an isolated fixed point, present multiple fixed points with their associated basins of attraction or even exhibits an oscillatory periodic attractor. All along this paper we will be primarily interested in a parameters setting that yields bistability.  As illustrated  in Figure \ref{fig1}, the system possesses  two attractive fixed points (green dots), which are
  separated by an unstable saddle point (not displayed). The (red) arrows stand for the stable and unstable manifolds associated to the saddle point. The stable direction marks the transition between the adjacent basins of attractions of the two linearly stable fixed points. Deterministic WC trajectories are asymptotically attracted towards one of the stable equilibria, depending on their initial condition. More interesting is instead the dynamics of the WC model in its stochastic representation:  finite size, hence endogenous, noise drives seemingly erratic transitions between the two, deterministically stable, fixed points. This picture is exemplified in  Figure \ref{fig2}, where the time evolution of species $x$ is plotted. This is an individual trajectory obtained upon integration of the Markov WC model via the usual Gillespie Monte Carlo scheme.

\begin{figure}
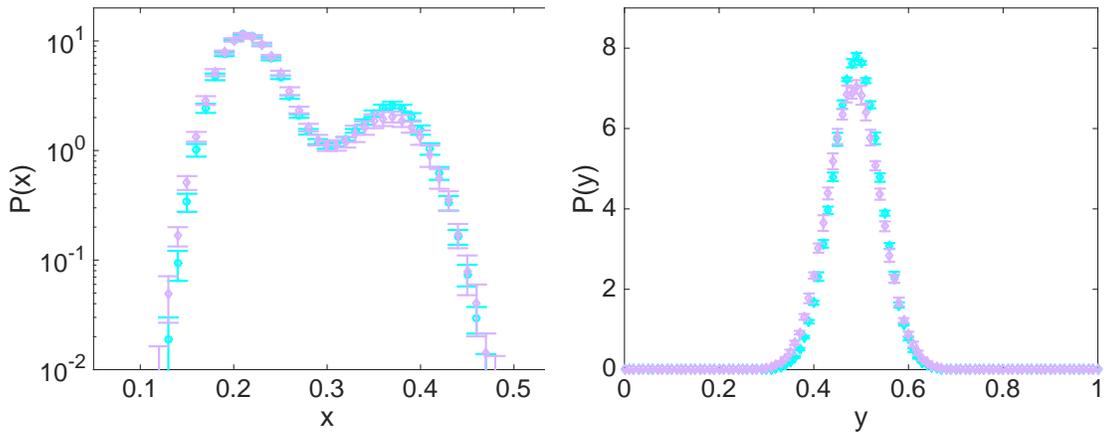

 \centering
 \begin{tabular}{cc}
 \includegraphics[scale=0.4]{figure3.eps} & \includegraphics[scale=0.4]{figure4.eps}
 \end{tabular}
 \caption{Left panel: the distribution $P(x)$ is plotted versus $x$ in semi-logarithmic scale. The (cyan) circles refer to the exact, Gillespie based, stochastic simulation. The error bars are computed by averaging over $10$ independent realizations. The (light purple) diamonds are obtained by integrating numerically the approximate Langevin equations (\ref{mult_lang}) via a standard Euler-Maruyama scheme \cite{euler_maruyama} and averaging over $10$ independent realizations. The agreement between the exact and approximate description is satisfying.  Right panel: $P(y)$ is plotted versus $y$. The choice of symbols (and colors) follows that of the left panel. At this level of noise, the distribution $P(y)$ appears single peaked, since the stable fixed points of the deterministic dynamics share  similar values of $y$ (see Figure  \ref{fig1}). Parameters are set as in Figure \ref{fig1}}.
 \label{fig3}
 \end{figure}

To test the adequacy of the approximate Langevin description (\ref{mult_lang}), we conducted a series of numerical simulations. The distribution $P(x)$ (resp. $P(y)$) as obtained via the Gillespie \cite{gillespie} algorithm, hence drawn in accordance with the exact governing master equation, are displayed in Figure \ref{fig3}, with (cyan) symbols. The homologous quantities computed via a direct integration of the Langevin equation is depicted with (light purple) diamonds. The agreement is satisfying and points to the validity of the approximate Langevin picture. The multiplicative component of the noise is essential to reach this level of correspondence at moderate system sizes. Notice that $P(x)$ shows the typical bimodal distribution, signature of a bistable dynamics.
Conversely, at this noise level the two bumps of $P(y)$ merge together in a unique isolated crest.

Starting from this preliminary analysis, the remaining part of this paper is devoted to analytically characterizing the above probability distributions in the region of bistability. We will in particular begin to explore a simplified setting, which neglects the excitatory feedback on the inhibitors population ($\omega_{EI}$). By operating in this context, one can significantly reduce the complexity of the model:  the two governing Langevin equations, as obtained under the Kramers-Moyal expansion, will be packed in just one stochastic differential equation for the excitatory species. Remarkably, the inhibitors will be shown to affect the amplitude of the multiplicative noise, and thus magnify the stochastic component of the excitatory dynamics.

\section{Analytic characterization of the distributions: the case $\omega_{IE}=0$.}
\label{sec2}

\begin{figure}
 \includegraphics[scale=0.4]{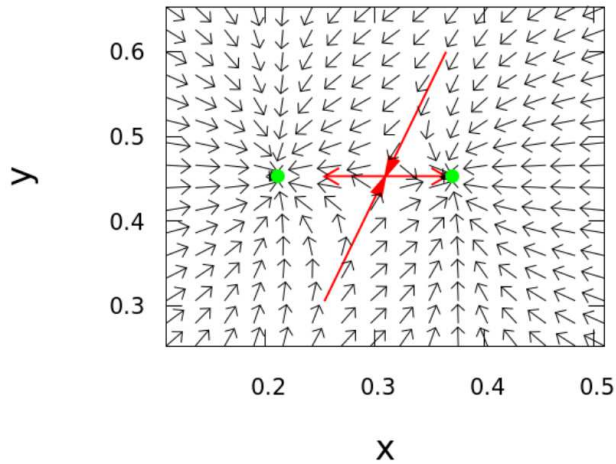}
 \caption{The phase space of the deterministic WC model is displayed. The (green) filled circles identify the location of the stable fixed points. Large (red) arrows stands for the
 stable and unstable manifolds of the saddle node (not shown). Tiny (black) arrows represents the vector field of the model.  The parameters are set as in Figure \ref{fig1},
 except for $\omega_{IE}=0$. }
 \label{fig4}
 \end{figure}

The aim of this section is to proceed further in the study of the fluctuating hydrodynamics of the WC model. More specifically, we will start from the Langevin equations (\ref{mult_lang}), which proved adequate versus numerical simulations of the underlying Markov process, to build up an approximate analytical solution for the distribution of of the stochastic variables $x$ and $y$.
To persecute this objective we shall specialize first on the simplified setting that is recovered when $\omega_{IE}=0$. In concrete, we will silence the retroaction of the activators on the inhibitors, while still  constraining the system to evolve in the region of deterministic bistability for the excitatory species.  The concentration of the inhibitors will instead converge to a stationary stable fixed point, in the mean field limit. As anticipated above, this is a definite simplification which enables us to reduce the system to just one Langevin equation for the excitatory species, the
multiplicative component of the noise being modified so to reflect the action of the inhibitory neurons. In Figure  \ref{fig4} the phase space portrait of the system as obtained under these operating condition is outlined. The two stable fixed points (represented as green filled circles) lay horizontally, since, by construction,  they display an identical value of inhibitory concentration.

For convenience, we write the Langevin equations (\ref{mult_lang}) in the compact form:

\begin{eqnarray}
\dot{x}&=&A(x,y)+\frac{1}{\sqrt{N}}\sqrt{B(x,y)}\eta_1 \\ \nonumber
\dot{y}&=&\frac{1}{\gamma} \left( C(y)+\frac{1}{\sqrt{N}} \sqrt{D(y)}\eta_2 \right)
\label{mult_lang1}
\end{eqnarray}

with:

 \begin{eqnarray}
A(x,y) &=& -\alpha_E x +(1-x)f_E(x,y) \\ \nonumber
B(x,y) &=& \alpha_E x+(1-x)f_E(x,y) \\ \nonumber
C(y) &=& -\alpha_I y +(1-y)f_I(y) \\ \nonumber
D(y) &=& \alpha_I y+(1-y)f_I(y)
\end{eqnarray}

and where $C(\cdot)$ and $D(\cdot)$ are just function of $y$, being  $\omega_{IE}=0$. To proceed in the analysis we take advantage of the fact that $y$ is a peaked distribution, under the chosen operating conditions. More precisely, we set to approximate the variable $y$ as:

\begin{equation}
\label{expansion}
y \simeq \bar{y}+u
\end{equation}

where $\bar{y}$ solves the mean field equilibrium condition $C(\bar{y})=0$ and $u$ encodes for the stochastic contribution. This latter quantity can be statistically characterized via a linear noise approximation, which amounts to inserting ansatz (\ref{expansion}) in the second of equations (\ref{mult_lang1}), expanding for small $u$ and arresting the perturbative calculation at the first
order. In formulae one readily gets:

\begin{equation}\label{langevin_fluctuations}
\dot{u}=\frac{1}{\gamma}\left( \partial_y C|_{\bar{y}} u +\sqrt{\frac{1}{N}}\sqrt{D(\bar{y}})\eta_2 \right)
\end{equation}

where $<\eta_2(t)\eta_2(t')> = \delta (t-t')$. This is a linear Langevin equation that can be solved to yield
$u(t) = u(0) \exp ( - t/t_c ) + \int^t_0 \exp (-(t-t')/t_c) \Gamma(t') \mathrm{d}t'$, where
$t_c = -\gamma/ \partial_yC|_{\bar{y}}$. Assuming the correlation time to decay sufficiently fast (hence for $t_c$ sufficiently small), it is legitimate to approximate $u$ as a Gaussian delta correlated noise with variance:

\begin{equation}
\sigma^2=-\frac{D(\bar{y})}{2N\gamma \partial_{y}C|_{\bar{y}}}
\end{equation}

Under the linear noise approximation, $P(y)$,  the probability distribution of inhibitory neurons, takes therefore the compact form:

\begin{equation}
\label{distry}
P(y)=\mathcal{N}\exp \left(- \frac{1}{2\sigma^2}\left(y -\bar{y} \right)^2\right)
\end{equation}

where $\mathcal{N}$ stands for a proper normalization constant. Let us now return to the first of equations (\ref{mult_lang1}) and assume, in light of the above,
$y \simeq \bar{y}+\sigma \eta_3$, where  $\eta_3$ is a normally distributed random variable, with zero mean and variance equal to one. Expanding the Langevin equation for $x$
at the first oder in the stochastic perturbation $v$ one obtains:

\begin{equation}
\dot{x}=A(x,\bar{y})+\partial_{y}A|_{\bar{y}} \sigma \eta_3+\frac{1}{\sqrt{N}}\sqrt{B(x,\bar{y})}\eta_1
\end{equation}

This is a non linear stochastic equation with two distinct noise sources, which can be readily combined by using the sum rule for Gaussian variables to yield the more compact expression:

\begin{equation}
\label{LangEqCombined}
\dot{x}=A(x,\bar{y})+ \frac{1}{N} \left(\sqrt{ -\frac{D(\bar{y})}{2 \gamma \partial_{y}C|_{\bar{y}}} \left(\partial_{y}A|_{\bar{y}} \right)^2 + B(x,\bar{y})} \right)\xi
\end{equation}

where $\xi$ is a normalized Gaussian white noise. The dynamics of the activator species is therefore ruled by a non linear Langevin equation, endowed with a multiplicative noise term, whose amplitude encodes, at this order of approximation, for the inhibitors contribution. Interestingly, the strength of the noise is modulated by $\gamma$.  Lowering the ratio $M/N$, which in turn signifies reducing the relative weight of inhibitory over excitatory neurons, enhances the degree of effective stochasticity.

Given the Langevin equation (\ref{LangEqCombined}), it is possible to write down an equivalent Fokker-Planck equation. This latter takes the form:

\begin{equation}
\frac{\partial P(x,t)}{\partial t}=-\frac{\partial}{\partial x}\left[ A(x,\bar{y})P(x,t)\right]+\frac{1}{2}\frac{\partial^2}{\partial x^2}\left[ \Gamma(x) P(x,t) \right]
\end{equation}

where:
\begin{equation}
\Gamma(x)=-\frac{D(\bar{y})}{2 \gamma \partial_{y}C|_{\bar{y}}} \left(\partial_{y}A|_{\bar{y}} \right)^2 + B(x,\bar{y})
\end{equation}

Based on the above, one can immediately calculate $P(x)$ the stationary probability distribution for species $x$, as \cite{gardiner}:

\begin{equation}
\label{distrx}
P(x)=\frac{\mathcal{N}}{\Gamma(x)}\exp \left( 2 \int^{x}_0 \frac{A(x',\bar{y})}{\Gamma(x')} \mathrm{d}x' \right)
\end{equation}

Equations (\ref{distry}) and (\ref{distrx}) constitute a theoretical approximation for the probability distributions of species $x$ and $y$, as determined via the governing stochastic model.
Recall however that the above solution applies to a rather specific parameter settings: $\omega_{IE}$ is in fact set to zero to silence
the retroaction of the activators on the inhibitors. Furthermore, we required the deterministic equation for $y$ to return a unique fixed point, an assumption that makes the linear noise approximation for the inhibitory species viable. Conversely, the population of excitatory neurons can execute a bistable dynamics, that, as we shall show hereafter, we will appropriately describe in terms of equation  (\ref{distrx}). The analysis will be extended in the forthcoming section to the case where a small modulation $\omega_{IE}$ is allowed for.

\begin{figure}
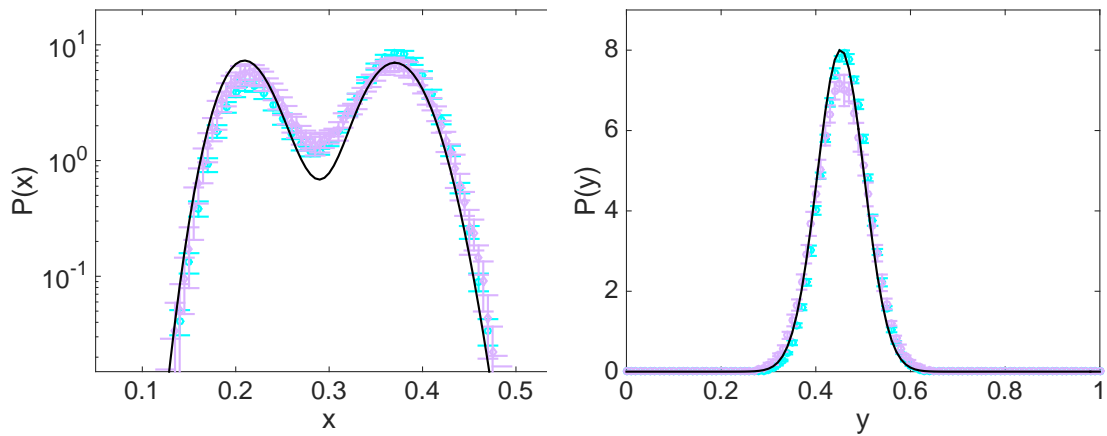

 \centering
 \begin{tabular}{cc}
 \includegraphics[scale=0.4]{figure6.eps} & \includegraphics[scale=0.4]{figure7.eps}
 \end{tabular}
 \caption{The distributions $P(x)$ (left panel) and $P(y)$ (right panel) are depicted.
 The (cyan) circles stand for the exact Gillespie stochastic simulation. The (light purple) diamonds refer to the numerical solution of the Langevin equations (\ref{mult_lang}), via a straightforward implementation of the Euler-Maruyama scheme. In both cases the error bars are computed by averaging over $10$ independent realizations. The solid lines are the theoretical predictions computed after equations (\ref{distry}) and (\ref{distrx}). Parameters are set as in Figure \ref{fig4}}.
 \label{fig5}
 \end{figure}

The remaining part of this Section is devoted to testing the predictive ability of expressions (\ref{distry}) and (\ref{distrx}), as derived above. In Figure \ref{fig5} the numerically determined
distributions are plotted: in both panels, (cyan) circles refer to the exact Master equation and (light purple) diamonds stand for a direct numerical integration of the Langevin equations (\ref{mult_lang1}). The parameters have been set as in Figure \ref{fig4}: species $y$ admits therefore a unique fixed point, while two stationary stable equilibria are found for $x$.  The distribution
$P(y)$ (right panel of Figure \ref{fig5}) displays the expected bell shaped profile, the peak being located in  $\bar{y}$. The solid line refers to the analytical solution  (\ref{distry}) which appears to correctly interpolate the recorded numerical profiles. More interesting is the distribution $P(x)$, as reported in the left panel of Figure \ref{fig5}: in this case, a bimodal profile is found, which
reflects the simultaneous presence of two competing equilibria for the associated deterministic dynamics. The theoretical profile traced after equation (\ref{distrx}) (solid line) agrees with the simulated data, at a satisfying degree of accuracy.

\begin{figure}
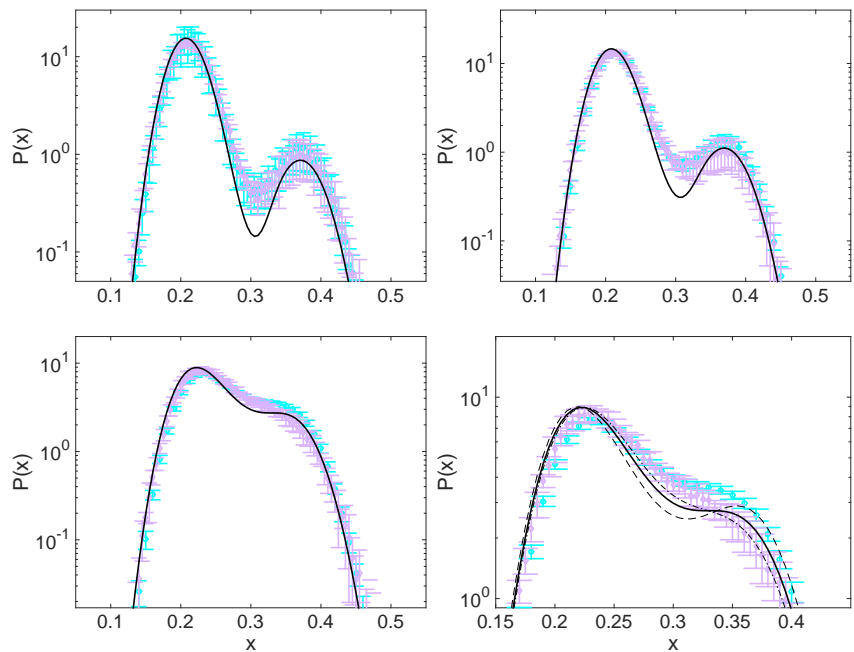

 \centering
 \begin{tabular}{cc}
 \includegraphics[scale=0.3]{figure8.eps} & \includegraphics[scale=0.3]{figure9.eps} \\
  \includegraphics[scale=0.3]{figure10.eps}  & \includegraphics[scale=0.3]{figure10b.eps}
 \end{tabular}
 \caption{The distributions $P(x)$ is plotted for different choices of the gain parameter $\beta_E$. The two upper panels refer to
 $\beta_E=4.5$ and $\beta_E=3.7$. In both cases the system falls in the region of deterministic bistability. The lower panels refer instead to
 $\beta_E=2$. In this case the deterministic bistability is lost, and the mean field dynamics for the species $x$ predicts the existence of an isolated stationary stable
 fixed point. The other parameters are set as in Figure  \ref{fig4}. Symbols stand for the simulations (see caption of Figure \ref{fig5} for details on the selected
 graphic outline). The solid line stands for the theoretical prediction based on formula (\ref{distrx}). Dot-dashed and dashed curves are traced according to  (\ref{distrx}) for, respectively,
 $\gamma=1$ and $\gamma=0.1$. }
 \label{fig6}
 \end{figure}

In Figure \ref{fig6} the comparison between simulated and theoretical $P(x)$ is drawn for different choices of the parameters $\beta_E$. The data reported in the upper panels of
Figure \ref{fig6}  are obtained for a choice $\beta_E$ that positions the system inside the region of $x$ bistability. The correspondence between theory and simulations is
again satisfying.  The lower left plot refers instead to a choice of $\beta_E$ that falls outside the region of bistability: in this case the deterministic model predicts the presence
of an isolated stable fixed point, characterized by a low level of activity. The distribution $P(x)$ extends however to large $x$, the small rightmost bump being a relic of the lost
bistability. By modulating $\gamma$, one can eventually make the second bump more pronounced, seeding a veritable noise induced bistability, which has no immediate counterpart in its corresponding deterministic framework.

\section{Extending the analysis to the general setting $\omega_{IE} \ne 0$.}
\label{sec3}

The purpose of this section is to discuss a possible extension of the analysis to the relevant setting $\omega_{IE} \ne 0$. This amounts to assume a
retroactive feedback of excitatory neurons on the population of inhibitors. The approach that we shall present hereafter implements a perturbative iterative scheme, which implies
dealing with a weak coupling, or stated differently, a sufficiently small value of $\omega_{IE}$. The method will prove however adequate for relatively large $\omega_{IE}$ amount, as for instance assumed in Figure \ref{fig1}.

The key idea is  to treat the additional contribution stemming from $\omega_{IE}$ as a perturbation. As a first step, one needs to preliminary inspect the null unperturbed scenario which corresponds to setting $\omega_{IE}=0$. To this end one can follow step by step the approach outlined in the preceding section and eventually obtain a probability distribution function for $x$. We will hereafter denote $P_0(x)$ the distribution obtained under this limiting condition, namely for $\omega_{IE}=0$.

In general  ($\omega_{IE} \ne 0$) the stochastic dynamics of $y$ is ruled by the following non linear Langevin equation:

\begin{equation}
\dot{y} = \frac{1}{\gamma} \left( C(z,y)+\frac{1}{\sqrt{N}}\sqrt{D(z,y)}\eta_2 \right)
\label{mult_lang2}
\end{equation}

where:

\begin{eqnarray}
C(z,y) &=& -\alpha_I y +(1-y)f_I(s_I) \\ \nonumber
D(z,y) &=& \alpha_I y+(1-y)f_I(s_I)
\end{eqnarray}

and $s_I=z-\omega_{EI}y +h_I$ with $z=\omega_{IE} x <<1$. By exploiting the fact that $\omega_{IE}$,  hence $z$, is small one can Taylor expand the function $C(z,y)$ in $z$ to eventually obtain:

\begin{equation}
C(z,y) \simeq C(0,y) +  \omega_{IE} x \frac{\partial s_I}{\partial z} \frac{\partial C}{\partial s_I} \mid_{0,y}+ O( \omega_{IE}^2)
\end{equation}

Notice that $\frac{\partial s_I}{\partial z}$ is the gain parameter, identically equal to $\beta_I$ for the family of non linear function $f_I(\cdot)$ here considered (see caption of Figure \ref{fig1}).
In equation (\ref{mult_lang2}) the function $D(z,y)$ is rescaled by $1/\sqrt{N}$, a small factor that encodes for the system size. For this reason,  we set as a first approximation  $D(z,y) \simeq D(0,y)$ and consequently drop all terms that combine  $\omega_{IE}$ and $1/\sqrt{N}$. Summing up, for small $\omega_{IE}$, the above equation  (\ref{mult_lang2}) can be cast in the
approximated form:

\begin{equation}
\dot{y} = \frac{1}{\gamma} \left( C(0,y) +  \omega_{IE} x \beta_I \frac{\partial C}{\partial s_I}\mid_{0,y} +\frac{1}{\sqrt{N}}\sqrt{D(0,y)}\eta_2 \right)
\label{mult_lang3}
\end{equation}

Notice that the deterministic term in equation (\ref{mult_lang3}) depends linearly on $x$, a non trivial modification which restores the bidirectional coupling with the homologous Langevin equation for species $x$. In principle, this fact prevents us from following the strategy adopted in the preceding Section. The assumed bistability on $x$ reflects in fact on species $y$, which also faces the simultaneous coexistence of two attracting equilibria. The linear noise approximation hence breaks down, as it is not possible to identify a unique fixed point of $y$ that would act as pivotal point for the linear noise expansion to be carried out. On the other hand, for $\omega_{IE}$ sufficiently small,  the two fixed points associated to species $y$ are evidently close. Then, it seems plausible to hypothesize that the two peaks that should correspondingly materialize in $P(y)$ would merge together, giving rise to a single bell shaped profile. Clearly,  the endogenous noise should be large enough (hence the system size sufficiently small) for the coalescence of the peaks to take place. This is indeed the scenario depicted in Figures \ref{fig1} and \ref{fig2}
(left panel) and which ultimately inspires the analysis to which the remaining part of this Section is entirely devoted.

\begin{figure}
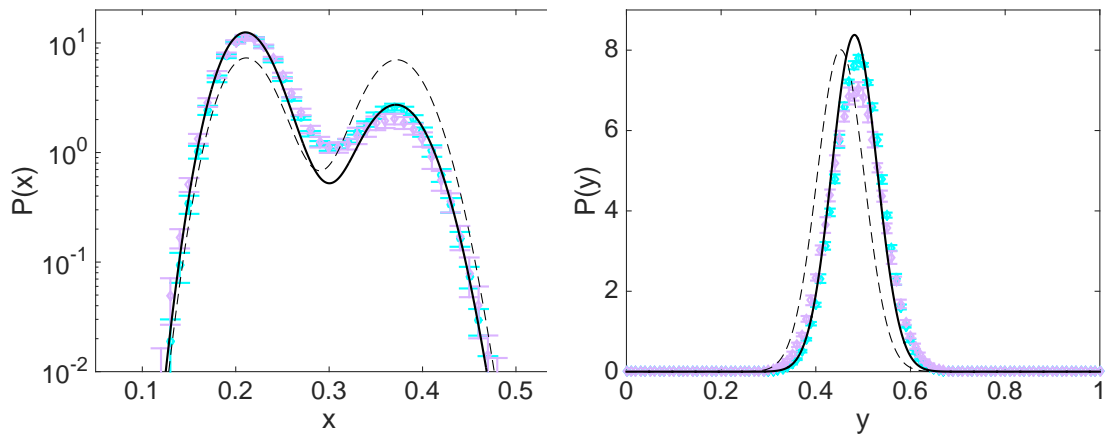

 \centering
 \begin{tabular}{cc}
 \includegraphics[scale=0.4]{figure11.eps} & \includegraphics[scale=0.4]{figure12.eps}
 \end{tabular}
 \caption{The distributions $P(x)$ (left panel) and $P(y)$ (right panel) are represented.
 The (cyan) circles refer to the stochastic Gillepsie based simulations. The (light purple) diamonds are obtained upon integration of the non linear Langevin equations (\ref{mult_lang}). In both cases the error bars are computed by averaging over $10$ independent realizations. The dashed lines represent the initial theoretical guess ($\omega_{IE}=0$). The solid lines stand for the solution obtained after three consecutive iterations of the method described in the main body of the paper. Parameters are set as Figures \ref{fig1} and  \ref{fig2}. In particular, $\omega_{IE}=0.5$.}
 \label{fig7}
 \end{figure}

To proceed further we propose to replace the time dependent factor $x$ appearing in equation (\ref{mult_lang3}) with $\bar{x}_0 = \int x P_0(x) dx$, a constant approximate quantity which follows the determination of the distribution function $P_0(x)$, as obtained in the limiting setting  $\omega_{IE}=0$.  The fixed point $\bar{y}_1$ of equation (\ref{mult_lang3}) should therefore match the condition $C(0,\bar{y}_1) +  \omega_{IE} \bar{x}_0 \beta_I \frac{\partial C}{\partial s_I}\mid_{0,\bar{y}_1}=0$. Here,  a sole fixed point is expected, the distribution $P(y)$ being, by assumption, single peaked. To characterize the fluctuations associated to the stochastic variable $y$, we perform a linearization of (\ref{mult_lang3}) around the fixed point $\bar{y}_1$. More explicitly we posit
$y=\bar{y}_1+u$,  and carry out a Taylor expansion in $u$, the supposedly small stochastic perturbation. A straightforward mathematical manipulation yields the linear stochastic equation:

\begin{equation}
\dot{u} = \frac{1}{\gamma} \left[ \left(\frac{\partial C}{\partial s_I }\right) \mid_{0,\bar{y}_1} + \left(\frac{\partial^2 C}{\partial y  \partial s_I } \right) \mid_{0,\bar{y}_1} \right] u+\frac{1}{\gamma \sqrt{N}}\sqrt{D(0,\bar{y}_1)} \eta_2
\end{equation}

where in the last term we have deliberately dropped the linear contribution in $u$.  As remarked earlier, in fact, the amplitude of the stochastic term
is already scaled by $1/\sqrt{N}$, a small factor that senses the size of the scrutinized system. Proceeding in analogy with
the above, we approximate $u$ as a Gaussian delta correlated noise of variance:

\begin{equation}
\label{sigma_1}
\sigma_1^2=\left( \frac{-D(0,\bar{y}_1)}{2 \gamma N  C'} \frac{1}{1+C^{''}/C^{'}}  \right)
\end{equation}

where $C'=\left( \frac{\partial C}{\partial s_I }\right) \mid_{0,\bar{y}_1} $ and $C^{''}=\left( \frac{\partial^2 C}{\partial y  \partial s_I } \right)\mid_{0,\bar{y}_1}$. Summing up, at this order of approximation, the stochastic variables $y$ follow a Gaussian distribution $P_1(y)$ of mean $\bar{y}_1$ and variance $\sigma_1^2$. Building on this observation we can proceed as illustrated in the preceding Section to eventually obtain an updated expression for the distribution of $x$, namely $P_1(x)$. The reasoning can be iterated further, by exploiting $P_1(x)$ to calculate a novel estimate
$\bar{x}_2$, which would allow in turn to self-consistently calculate $P_2(x)$ and $P_2(y)$, the updated distribution of, respectively, $x$ and $y$. The convergence of the scheme is not guaranteed a priori, nor proven on solid mathematical grounds, but a posteriori confirmed by direct numerical inspection.  In Figure \ref{fig7} the distributions computed according to the above procedure are confronted to direct simulations assuming the parameters setting of  Figures \ref{fig1} and  \ref{fig2}. The zeroth order of approximation corresponding to the limiting condition
$\omega_{IE}=0$ is depicted with a dashed line. The solid line refer to the prediction obtained after three successive iterations of the outlined procedure. The method converges steadily, but the improvements are already remarkable after the first round of iteration.

\section{Conclusion}

Endogenous noise stemming from finite size fluctuations can play a role of paramount importance in shaping the self-emerging macroscopic dynamics of a multi-species model. Demographic
perturbations resulting from the inherent discreteness of the inspected medium can be rationalized by resorting to a linear noise treatment, which proves perfectly adequate when the underlying
deterministic dynamics admits a trivial fixed point. In general, non linearities are present and may eventually yield more complex dynamical behaviors, including multistability. In this case the linear noise machinery can be punctually employed to resolve the local dynamics of the system around each of the mutually competing equilibria. To quantify the statistics of the rare events that materialize in the transitions from one attractor of the deterministic dynamics to another, and so reconstruct the underlying stationary probability distributions, requires extending beyond the domain of application of the linear noise approximation.

Working in this framework, we have here considered a stochastic version of the Wilson Cowan (WC) model, which accomodates for a finite population of excitatory
 and inhibitory neurons. The model assumes a finite carrying capacity, the population census summing up to a constant. The relative ratio of inhibitory over excitatory neurons acts as a control parameter of the model.  The master equation that rules the dynamics of the stochastic WC model is expanded in powers of the inverse population size:
a fluctuating hydrodynamics picture is  consequently obtained, which consists in a coupled pair of non linear Langevin equations, with multiplicative noise. This latter equations are solved numerically and shown to return a statistical description of the stochastic dynamics, which adheres quantitatively with that displayed by the exact Markov model. Our analysis is specifically targeted to the region of bistability, a phenomenon that can a priori interest both the interacting families. To gain analytical insight into the investigated process we however specialize on a simplified scenario, assigning the parameters so to have just one stationary stable equilibrium for the population of inhibitors. The excitatory neurons can instead undergo bistable behavior.  Working in this setting, we operate a substantial reduction in complexity. In particular, the dynamics of the excitatory species is ruled by an independent, non linear Langevin equation:
the indirect imprint of the inhibitors is reflected in a modified noise term, while the deterministic contribution is evaluated at the fixed point of species $y$.  The population of inhibitors can be ideally schematized as an additional noise source. This latter contribution gets more pronounced the smaller the relative ratio of inhibitory versus excitatory neurons, broadening in turn the region where a bistable-like behavior is  detected. Analytical expressions for the distributions of, respectively, excitatory and inhibitory neurons are obtained, which show a satisfying degree of agreement with the simulated data. The methods are subsequently generalized to the relevant setting where both species display multiple fixed points. An iterative scheme is in fact devised that builds on the zeroth order approximation described above, and allows us to explain the modifications displayed in the recorded probability distribution function, when a modest feedback of the excitatory on inhibitory neurons is accommodated for. In conclusion we have contributed to the understanding of the stochastic WC model, focusing on a dynamical regime for which the linear noise approximation is manifestly inadequate. The non linear Langevin equations which represent the fluctuating hydrodynamic limit of the original model, can be analytically characterized under suitable operating conditions, so contributing to shed light on the subtle dynamical interplay between distinct families of interacting neurons.

\begin{acknowledgments}
This work has been supported by the H2020-MSCA-ITN-2015 project COSMOS  642563.
\end{acknowledgments}

 \end{document}